\documentclass[
    ,final            % use final for the camera ready runs
  ]
  {aipproc}

\layoutstyle{8x11single}
\usepackage{graphicx}

\newcommand{\GeV}{\makebox{ GeV}}
\newcommand{\beq}{\begin{equation}}
\newcommand{\enq}{\end{equation}}
\newcommand{\beqa}{\begin{eqnarray}}
\newcommand{\beqast}{\begin{eqnarray*}}
\newcommand{\enqa}{\end{eqnarray}}
\newcommand{\enqast}{\end{eqnarray*}}

\def\GeV{\nobreak\,\mbox{GeV}}

\begin{document}

\title{Elastic Amplitudes and Observables in pp Scattering}

\author{A. Kendi Kohara}{
  email={kendi@if.ufrj.br}, 
address={Instituto de F\'{\i}sica, Universidade Federal do Rio de
 Janeiro, C.P. 68528, Rio de Janeiro 21945-970, RJ, Brazil  } }
\author{Erasmo Ferreira}{email={erasmo@if.ufrj.br}}  
 \author{Takeshi Kodama}{email={tkodama@if.ufrj.br} }

\begin{abstract}
 
Using a unified analytic representation for the elastic scattering
amplitudes of pp scattering valid for all high energy region, the behavior of
observables in the LHC collisions in the range $\sqrt{s}$ = 2.76 - 14 TeV is
discussed. Similarly to the case of 7 TeV data, the proposed amplitudes give
excellent description of the preliminary 8 TeV data. We discuss the expected
energy dependence of the observable quantities, and present predictions for
the experiments at 2.76, 13 and 14 TeV.

\end{abstract}

\maketitle

%%%%%%%%%%%%%%%%%%%%%%%%%%%%%%%%%%%%%%%%%%%%
%% MAINMATTER
%%%%%%%%%%%%%%%%%%%%%%%%%%%%%%%%%%%%%%%%%%%%

\section{General information and Data Analysis }

We establish explicitly disentangled real and imaginary
amplitudes for pp elastic scattering based on a QCD motivated model.
With impact parameter representation   $(s,\vec{b})$     and
its Fourier transform in $(s,\vec{q })$  space  both represented by 
simple analytical forms, 
we are able to control unitarity and dispersion relation constraints, and provide
geometric interpretation of the interaction range. 
The regularity obtained in the description of the data and the physical interpretation give reliability to the proposed amplitudes.

The amplitudes of  pp elastic scattering  
  originally constructed through profile functions in  $b$ 
 -space  are written  
\begin{equation}
\widetilde{T}_{K}(s,\vec{b})=\frac{\alpha _{K}}{2\beta _{K}}e^{-{b^{2}}/{%
4\beta _{K}}}+\lambda _{K}(s)\widetilde{\psi }_{K}(\gamma _{K}(s),b)~,  \label{b-AmplitudeN}
\end{equation}%
with the usual  Gaussian forms plus  the characteristic shape functions 
\begin{equation}
\widetilde{\psi }_{K}(s,b)=\frac{2e^{\gamma _{K}-\sqrt{\gamma _{K}^{2}+{b^{2}%
}/{a_{0}}}}}{a_{0}\sqrt{\gamma _{K}^{2}+{b^{2}}/{a_{0}}}}\Big[1-e^{\gamma
_{K}-\sqrt{\gamma _{K}^{2}+{b^{2}}/{a_{0}}}}\Big]~.  \label{Shape-b}
\end{equation}%
The label $K$ $=R,I$ indicates either the real or the imaginary part of the
complex amplitude.

For large $b$,  
corresponding to peripheral collisions, the amplitudes fall down with a
Yukawa-like tail, $ \sim (1/b)\exp(-b/b_{0})$,   
 that reflects effects of virtual partons (modified gluon field)  at
large distances in the Stochastic Vacuum Model  \cite{dosch}. 

The comparison with $d\sigma /dt$ data and determination of parameters 
are made with the amplitudes in $t$-space. The quantities $\Psi
_{K}(\gamma _{K}(s),t=-\vec{q}_{T}^{2})$ obtained by Fourier transform of
Eq. (\ref{b-AmplitudeN}) are written 
\begin{equation}
\label{t_space}
T_{K}^{N}(s,t)=\alpha _{K}(s)\mathrm{e}^{-\beta _{K}(s)|t|}+\lambda
_{K}(s)\Psi _{K}(\gamma _{K}(s),t) ~,   \label{TN-b1}   
\end{equation} 
and the shape functions converted to $t-$ space take the form 
 \begin{eqnarray}
 \Psi _{K}(\gamma _{K}(s),t)  
 =2~\mathrm{e}^{\gamma _{K}}~\bigg[{\frac{\mathrm{e}^{-\gamma _{K}\sqrt{%
 1+a_{0}|t|}}}{\sqrt{1+a_{0}|t|}}}-\mathrm{e}^{\gamma _{K}}~{\frac{e^{-\gamma
  _{K}\sqrt{4+a_{0}|t|}}}{\sqrt{4+a_{0}|t|}}}\bigg]~.    
   \label{psi_st}
  \end{eqnarray}
 
The complete analysis of elastic scattering 
 \cite{ferreira1, KEK_2013, KEK_2013b} 
 requires also   the contributions from the Coulomb interaction at small 
$|t|$ and from
perturbative 3-gluon exchange at large $|t|$. The fixed parameter $a_0=1.39~ \GeV^{-2}$ 
is given by the correlation length of the gluon condensate. All parameters have been 
determined as smooth functions of $s$  and the properties of the amplitudes 
(magnitudes, signs, zeros) have been described in detail \cite{KEK_2014b}.  
 
In our normalization the elastic differential and the total cross sections
are written 
\begin{eqnarray}  \label{Sigma_diff}
\frac{d\sigma(s,t)}{dt}= (\hbar c)^2[T_I^2(s,t)+ T_R^2(s,t)]  
= \frac{d\sigma^I(s,t)}{dt} + \frac{d\sigma^R(s,t)}{dt} ~ ~ ~ , ~ ~ ~
\sigma(s) = (\hbar c)^2~ 4\sqrt{\pi} ~ T_{I}^{N}(s,t=0)~.
\end{eqnarray}

 \subsection{Differential Cross Sections and Amplitudes 
           in the 1.8 to 14 TeV Range  \label{energy_range}   }

\begin{figure*}[b]
\includegraphics[width=7cm]{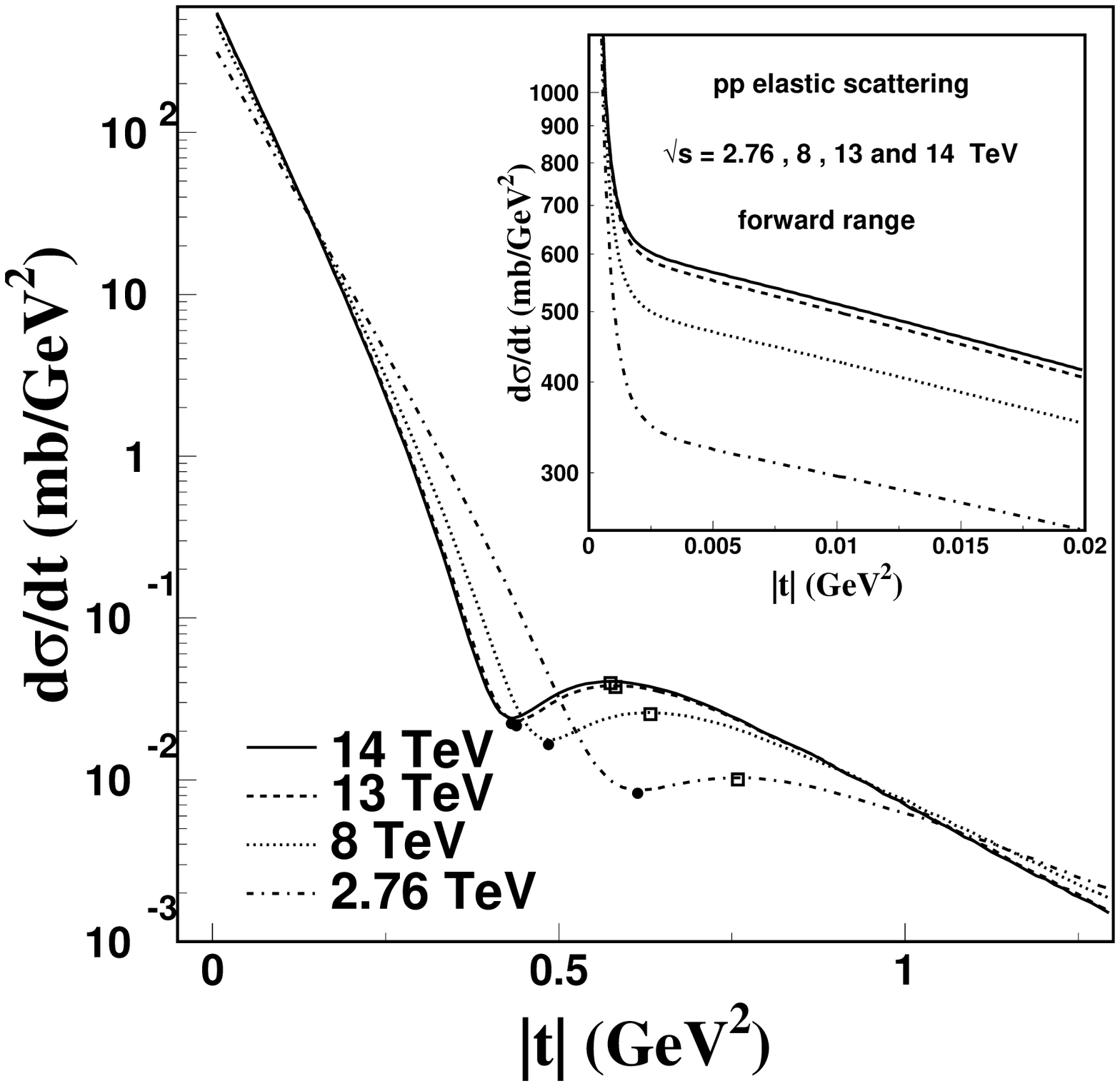}
\includegraphics[width=7cm]{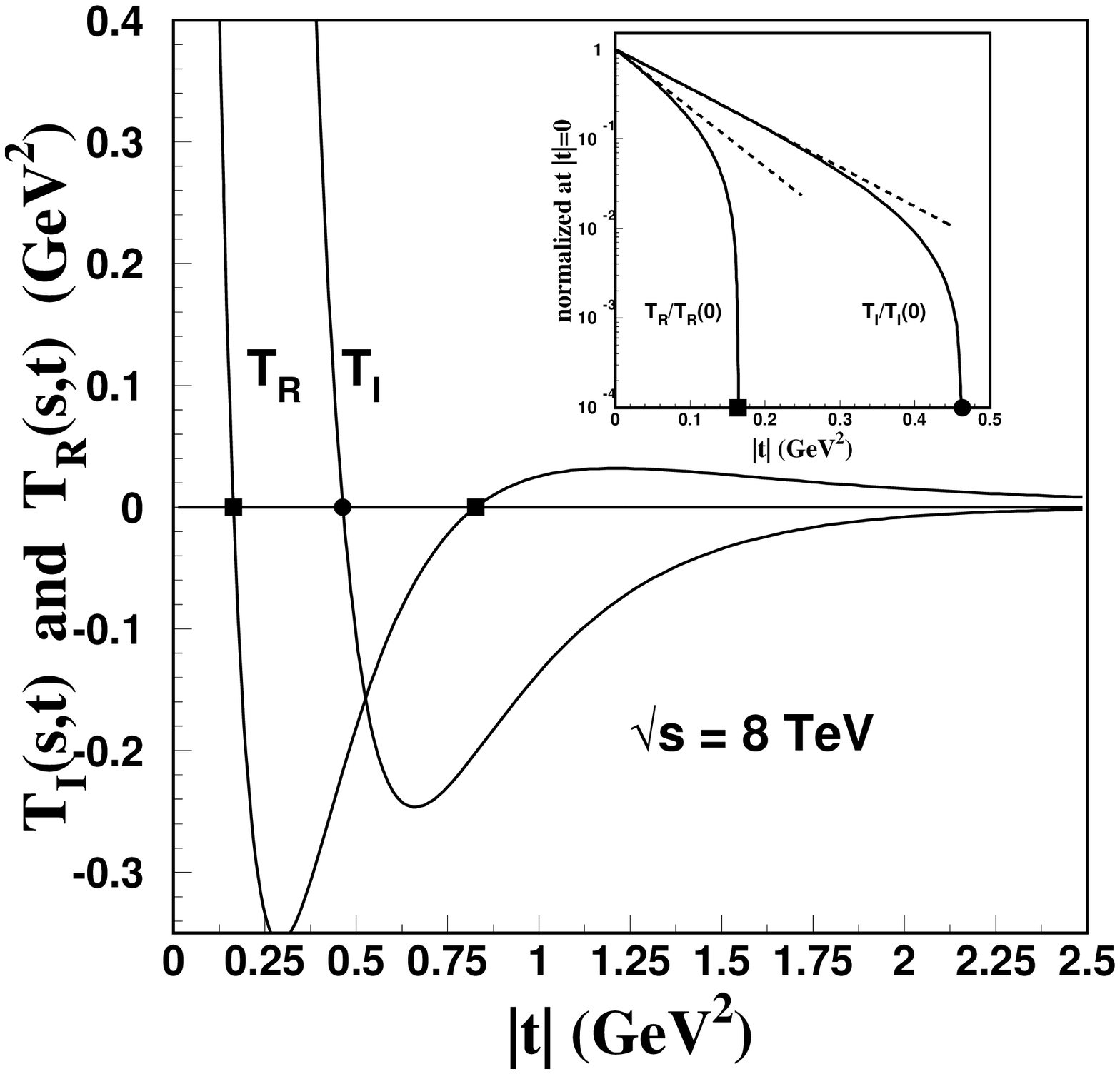} 
\caption{(LHS) -  Values of $d\protect\sigma /dt$ obtained for
energies of LHC experiments.  The positions of
dips and bump peaks, marked with dots and squares, 
 displace to the left as the energy increases, and can
be connected with straight lines. The inset shows the low $|t|$ range, with
Coulomb interaction effects included. 
(RHS) - Real  and imaginary parts of elastic pp scattering
amplitude at 8 TeV, as functions of $|t|$. The general behaviour is the same
for all energies, with one and two zeros respectively for the imaginary and
real parts. The behaviour for small $|t|$ is shown in the inset, indicating
the difference of slopes $B_R$ and $B_I$, and the deviations
of the exponential forms that occur as $|t|$ increases, each amplitude going
towards its zero. A second zero of the imaginary part occurs at much higher $%
|t|$. }
\label{pilhas-fig}
\end{figure*}

In Fig. \ref{pilhas-fig} we show the predictions of $d\sigma /dt$ for the
LHC energies 2.76 , 8 , 13 and 14 TeV.  
In the RHS  we use the energy $\sqrt{s}=8$ TeV as an
example to show the imaginary and real nuclear  amplitudes $T_I^N(s,t)$, $T_R^N(s,t)$
as functions of $|t|$ as predicted by Eq.(\ref{t_space},\ref{psi_st}). 
  The interplay of the imaginary and real amplitudes
at mid values of $|t|$ is responsible for the dip-bump structure of the
differential cross section, that was shown before \cite{KEK_2013} for $\sqrt{%
s}=7$ TeV. For $|t|\geq
1.5 \nobreak\,\mbox{GeV}^2$ the real part becomes dominant, with positive
sign. 

%  \section{ Comparison with Data and Predictions for 8 TeV  \label{data}}

Our description \cite{KEK_2013} of the elastic scattering data at 7 TeV from
the TOTEM Collaboration \cite{TOTEM_7} reproduces N=165 points in $d\sigma/dt
$ with an impressive squared average relative deviation $<\chi^2>=0.31$.
Characteristic quantities at this energy 
 are $\sigma=98.65$ mb, $\sigma_{\mathrm{el}}=25.39$
mb , $B=19.90$ GeV$^{-2}$, that compare extremely well with the values
 $\sigma=98.6 \pm 2.2$ mb, $\sigma_{%
\mathrm{el}}=25.4 \pm 1.1$ mb , $B=19.9 \pm 0.3$ GeV$^{-2}$ 
published by TOTEM.

\subsection{Preliminary data for $d\protect\sigma /dt$ at 8 TeV 
\label{dsigdt_8TeV}}
\begin{figure*}[b]
\includegraphics[width=7cm]{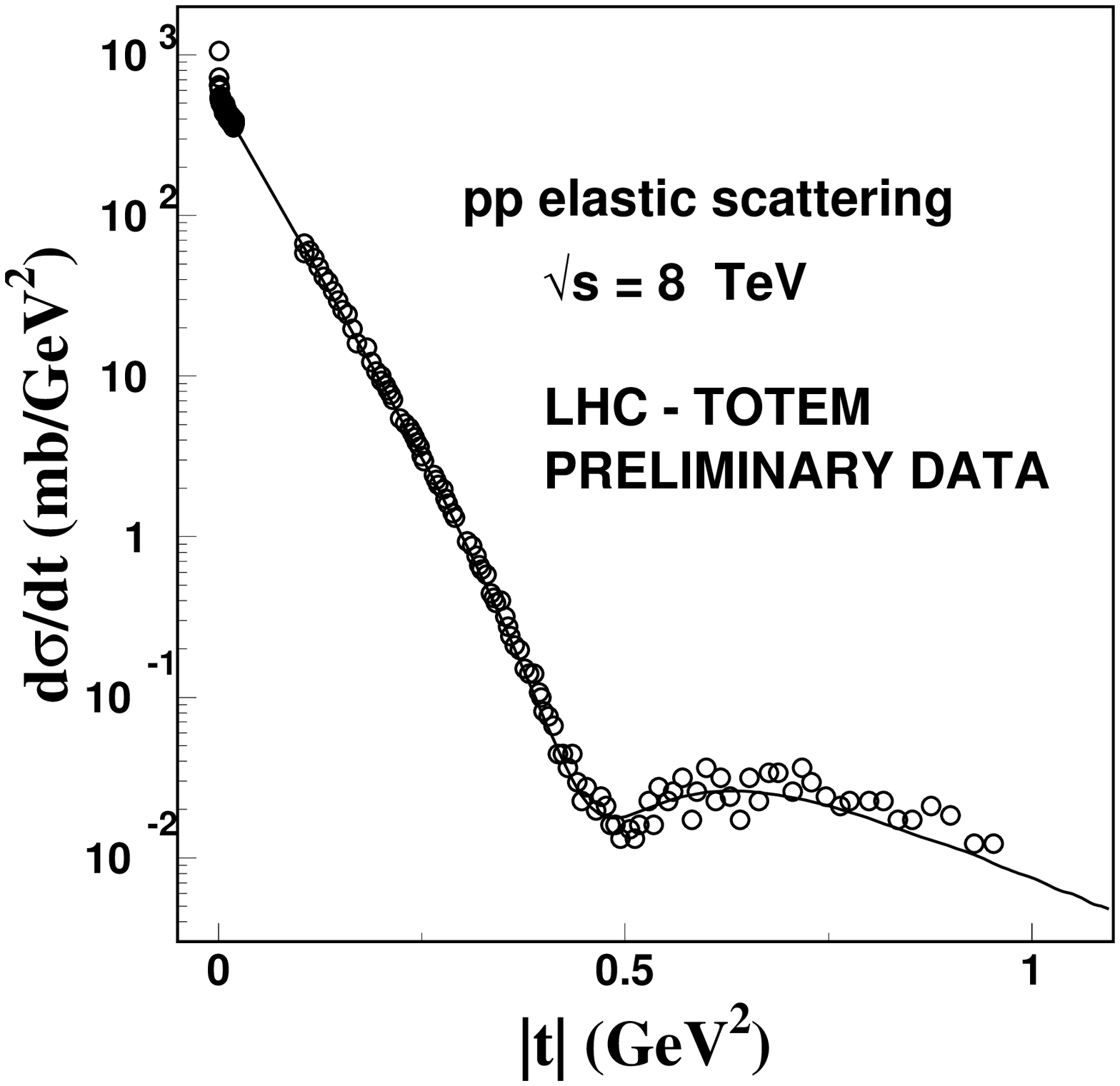} 
\includegraphics[width=7cm]{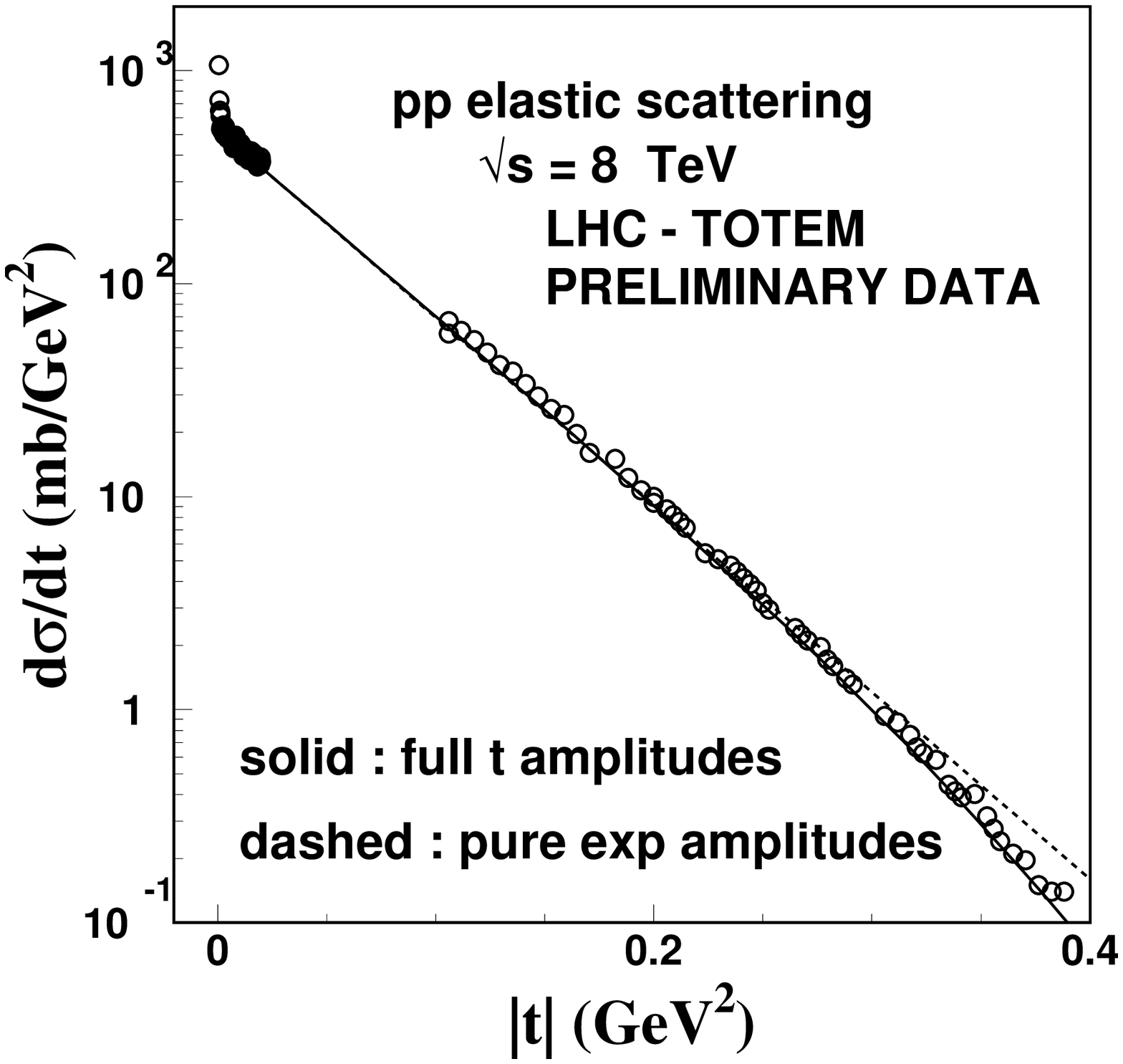}  
\caption{ Preliminary data at 8 TeV extracted by eye from presentation
slides of the TOTEM collaboration plotted together with our predicted
representation for $d\protect\sigma/dt$. 
 In the RHS we plot the forward range, with  the  dashed
line for pure exponential amplitudes. }
\label{data_8TeV-fig}
\end{figure*}
The preliminary TOTEM data of $d\sigma/dt$    
 at 8 TeV seem to be
regular enough for   our   analysis. The
  data discussed below are taken (reading by eye) from slides of
talks by members of the TOTEM Collaboration \cite{TOTEM_talks}.

As far as we can read from the presentation slides, we identify 212 data
points in three sets : 1) N=97 points in the forward interval $6\times
10^{-4} \leq |t| \leq 0.02$ GeV$^2$ ; 2) N=45 points in an intermediate
interval $0.08 \leq |t| \leq 0.3$ GeV$^2$ ; 3) N=70 points in a mid $|t|$
range $0.3 \leq |t| \leq 0.95$ GeV$^2$.
This information is transferred to the plot in Fig. \ref{data_8TeV-fig},
together with our calculation. The
  quality of the representation seems to be  equivalent to that of our
treatment of the 7 TeV data. 
     
Our values for $B_I$ and $B_R$    lead to the $d\sigma/dt$ effective slope 
$ B=[{B_I+\rho^2 B_R}]/[{1+\rho^2}] $ 
equal to $B=20.405 ~ \nobreak\,\mbox{GeV}^{-2}$.

Hopefully the analysis of the final TOTEM measurements will be more precise 
and will confirm the validity of our description for 8 TeV  and 
predictions for 13 and 14 TeV .

\subsection{ Inelastic and Total Cross Sections}

For the inelastic cross section we assume the difference $\sigma_{\mathrm{%
inel}}=\sigma-\sigma_{\mathrm{el}}$ and then we have  73.26 mb at 7 TeV..
Published values of the TOTEM Coll. using different methods are $73.15 \pm
1.26$ \cite{TOTEM_7}, $73.7 \pm 3.4$ \cite{TOTEM_7b} and $72.9 \pm 1.5$ \cite%
{TOTEM_7c}. ALICE Coll. \cite{ALICE} gives  $\sigma_{\mathrm{inel}}= 73.2
\pm 5.3 $ mb , and ATLAS Coll. $\sigma_{\mathrm{inel}}= 69.4 \pm 2.4 \pm 6.9 
$ mb \cite{ATLAS}. There are also  CMS results \cite{CMS} with 
  non-informed missing
contributions. In these measurements there are extrapolations  using Monte
Carlo models to include diffractive events of low mass. All these
results are compatible with our calculations.

A  measurement to be compared with our predictions is the $\sqrt{%
s}=2.76$ TeV value of ALICE Coll., that gives $\sigma_{\mathrm{inel}} = 62.8
\pm 4.2 $ mb , while we have the compatible value 63.11 mb.

The analysis of compatibility for the 1.8 TeV measurements of $\sigma_{%
\mathrm{inel}}$ by CDF and E811 in Fermilab  \cite{Klimenko} suggests the
value $ (1+\rho^2)\sigma_{\rm inel}=(60.3 \pm 2.3$ mb, that  with our $\rho$
value gives  $ \sigma_{\rm inel}=(59.1 \pm 2.3$ mb. We have 58.89 mb 
for 1.8 TeV, once more in very good agreement.

At 57 TeV the Auger Cosmic Ray experiment \cite{Auger}, 
using other models for  pp input,  obtains 
$ \sigma_{\mathrm{inel}} = 92 \pm 14.8 $ mb , while our extrapolation gives
101 mb. This measurement \cite{KEK_CR_2014} is discussed together with
other CR Extended Air Showers (EAS) experiments, using our inputs and a 
basic Glauber method to connect pp and p-air processes. Our
calculation reproduces well all CR data for p-air cross sections  
with energies  $\sqrt{s}$ (in the pp system)  up to 100 TeV.   
 
For 8 TeV we have predictions $\sigma=101.00$ mb , $\sigma_{\mathrm{el}%
}=26.18$ mb , $\sigma_{\mathrm{inel}}=74.82$ mb , $\sigma_{\mathrm{el}%
}/\sigma=0.26 $. The measurements by TOTEM \cite{TOTEM_8}
give for the same quantities $\sigma=101.7 \pm 2.9 $ mb , $\sigma_{\mathrm{el%
}}=27.1 \pm 1.4 $ mb, $\sigma_{\mathrm{inel}}=74.7 \pm 1.7$ mb, $\sigma_{%
\mathrm{el}}/\sigma=0.266\pm0.006$.

 All this information show that our formulae  $d\sigma/dt$ at 8 TeV and for the
energy dependence of $\sigma$ and $\sigma_{\mathrm{inel}}$ in pp scattering
work very well.

\subsection{ Remarks and Comments}

The proposed amplitudes have simple forms, evaluated with few  operations 
with elementary functions.
The shape of the dip-bump behavior results from a  delicate interplay 
of the imaginary and real amplitudes. All intervening quantities and 
  derived  properties present smooth energy dependences.
The zeros of the real and imaginary parts have very regular 
displacements, converging to   finite limits as the energy 
increases, showing remarkable connection with 
positions and heights of dips, bumps and  inflections in $d\sigma/dt$. 

The slopes $B_I$ and $B_R$ at the origin, with their characteristic 
difference in values, together with the ratio $\rho$,  are essential 
quantities  in the definition, through the unique analytical 
forms of the amplitudes, of the properties of the observed $d\sigma/dt$  
in the whole $t$ range. 
Their values are thus fixed with high accuracy. It is   
important that the slopes show quadratic dependence in $\log{s}$,
instead of the linear dependence suggested by Regge phenomenology. 

The integrated elastic cross sections are evaluated  in their 
separate parts, obtained from the real and imaginary amplitudes, and 
are also represented by simple parabolic forms in $\log{s}$. 

The properties of ratios (with respect to the total cross section)  
of slopes and of integrated elastic cross sections,  that 
tend to finite asymptotic  limits, show that the 
hypothesis of a black disk limit in the behaviour of the pp 
interaction  seems to be excluded by phenomenology.

Taking into account previous publications at 1.8 and 7 TeV, 
we obtain cross sections and amplitudes  at 
2.76, 8 , 13 and 14 TeV, with no free  numbers. Future data will test. 
 
We also discuss the geometrical interpretation of our amplitudes, showing 
that the effective interaction radius in $b$-space increases with the energy.
 Our amplitudes obey a geometric scaling in asymptotic energies, and
indicate that the profile function $d^{2}\sigma _{\mathrm{inel}}/d^{2}\vec{b}
$ tends to a universal (energy independent) function with respect to a
scaling variable, $x\sim b/\sqrt{\sigma }$.  This universal function exhibits 
a considerable diffused surface, indicating a   scenario different from 
the commonly accepted  black disk. At LHC energies, the saturation seems to start
(the central value of $d^{2}\sigma _{\mathrm{inel}}/d^{2}\vec{b}$ is almost
unity), but the asymptotic profile is still far and only  can be reached for 
$\sqrt{s}>10^{4} $ TeV. The connection between the diffused surface of long 
range and inelastic diffractive processes will be an interesting line of 
investigation.

We believe that our analytic representation of the
scattering amplitudes will serve as important guidance for the
future measurements in LHC,  and also for a theoretical understanding
 of the intermediate region of partonic saturation phenomena.

At this Diffraction 2014 conference related work on the black-disk 
behaviour \cite{menon} and on the determination of  pp amplitudes 
at LHC energies \cite{BSW,Selyugin}  were presented, showing that this 
is an important  field of research at the present.

\begin{theacknowledgments}
   The authors wish to thank the Brazilian agencies CNPq, PRONEX , CAPES and FAPERJ for
financial support.
\end{theacknowledgments}

\bibliographystyle{aipproc}



\bibliographystyle{aipproc}   % if natbib is available


\end{document}